\documentclass[prb,twocolumn,showpacs,superscriptaddress]{revtex4-1}

\usepackage{graphicx,amssymb,amsmath,color,bbm}

\begin{document}
\definecolor{darkgreen}{rgb}{0,0.5,0}

\newcommand{\jav}[1]{{#1}}

\title{Coupling, merging, and splitting Dirac points by electron-electron interaction}
\author{Bal\' azs D\' ora}

\email{dora@eik.bme.hu}

\affiliation{Department of Physics and BME-MTA Exotic  Quantum  Phases Research Group, Budapest University of Technology and Economics, Budafoki \'ut 8, 1111 Budapest, Hungary}

\author{Igor F. Herbut}

\affiliation{ Max-Planck-Institut f\"ur Physik Komplexer Systeme, N\"othnitzer Str. 38, 01187 Dresden, Germany}
\affiliation{ Department of Physics, Simon Fraser University, Burnaby, British Columbia, Canada V5A 1S6}

\author{R. Moessner}

\affiliation{ Max-Planck-Institut f\"ur Physik Komplexer Systeme, N\"othnitzer Str. 38, 01187 Dresden, Germany}

\date{\today}

\begin{abstract}
The manipulation and movement of Dirac points in the Brillouin zone by the electron-electron interaction is considered within leading order perturbation theory.
At the merging point, an infinitesimal
interaction is shown to cause opening of the gap or splitting of the Dirac points, depending on the inter- or intrasublattice nature of the merging and the sign of the interaction.
The topology of the spectrum
can therefore be efficiently changed by simply tuning the
interaction between particles, as opposed to the usual careful band structure engineering. This is illustrated around the merging transition of one, two, and three
dimensional Dirac-Weyl fermions. A simple
Weyl-like Hamiltonian that describes the quadratic band-crossing in three dimensions is also proposed, and its stability under interactions is addressed.
\end{abstract}


\maketitle

\section{Introduction}

The effects of many-body interactions on systems with Dirac dispersion have been long investigated. In one dimensional systems,
the Fermi-liquid breaks down and becomes replaced by the Luttinger liquid, with collective bosonic excitations.\cite{giamarchi}
In addition, the interactions can also alter or remove the low energy excitations completely,  and drive the system towards superconducting,
charge, or spin ordering instabilities. A prototypical example of such behavior is provided by carbon nanotubes. In two dimensions,
the Dirac cones in graphene and in topological insulators\cite{hasankane} have also attracted a lot of attention\cite{kotov}.
In particular, the effect of the long range Coulomb interaction,\cite{herjurvaf, mish, semenoffherbut, jurisic, rosensteininteraction}
especially important in suspended graphene samples, is found to significantly renormalize the Fermi velocity\cite{gonzalez1}, but
still not to lead to any  phase transitions. This has also been confirmed experimentally\cite{elias}. On the other hand, a sufficiently
strong short range component of the Coulomb interaction, such as the Hubbard $U$, for example, can indeed induce insulating
behaviour\cite{tosatti, paiva, herbut,hjr, meng, sorella, assaad}, although in the weak coupling limit such interactions are rather harmless.
In three dimensions, topological nodal semimetals\cite{balents, burkov, delplace} host Weyl fermions as elementary excitations, giving rise to topologically non-trivial gapless band structures, which are expected to be rather stable with respect to gap opening  due to various interactions.

Descendants of monolayer graphene exhibit other unique band structures as well, without any analogs in high energy physics, such as
the quadratic band crossing of AB stacked bilayer\cite{evcastro} or the cubic band crossing in ABC stacked trilayer graphene\cite{Guinea-Castro-Peres-2006,Min-MacDonald-2008}. While most of these peculiarities occur in conventional condensed matter systems as well,
the realization of some other of their further features require different settings. For example, by altering the uniform hopping integrals on honeycomb lattice, 
the two inequivalent Dirac cones are moved in the Brillouin zone, and can even be merged together. However, in graphene such merging would need an application of 
strains which are not physically accessible. Fortunately, engineering molecular graphene\cite{manoharan}, or using cold atoms\cite{tarruell} can lead to the observation of the desired effect \jav{of merging Dirac cones.}

The latter offers the versatility of not only changing the band structure and reaching the merging point in 2D, but also of tuning the interaction between the particles, or even changing their statistics, by loading different atoms into the optical lattice. Motivated by these exciting new possibilities, we have investigated the effect of interactions around the merging point of Dirac-Weyl systems in various dimensions.

\begin{figure}[h!]
\includegraphics[]{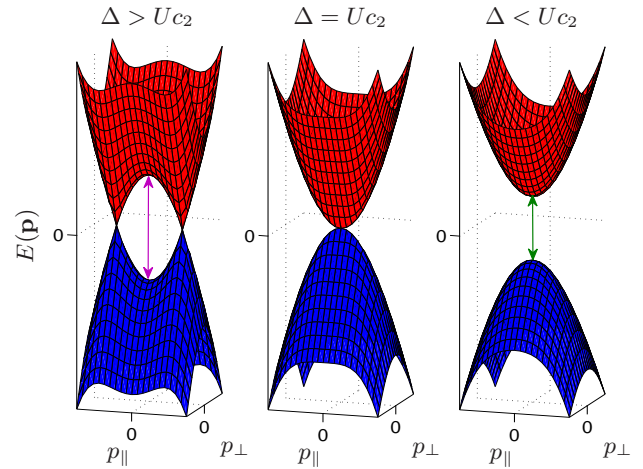}
\caption{The evolution of the spectrum as a function of $\Delta$ and the interaction $U$ for the 2D case. The spectrum varies as $\sim |p_\perp|$ around the zeros of $E(p)$.
The magenta arrow denotes the distance between the saddle points, $2(\Delta-Uc_2)$, and the green arrow stands for the gap, $2(Uc_2-\Delta)$, $c_2$ is a non-universal constant which we evaluate in the Hartree-Fock approximation. }
\label{specevol}
\end{figure}

In order to have Dirac points, we need an effective two-band model containing two species of fermions, coming from e.g. the two sublattices, as is the case for the honeycomb lattice\cite{castro07,tarruell}.
Close to the merging point, the inter-particle  interaction renormalizes the distance between the nearby Dirac points already at the Hartree-Fock level.
This is controlled by the term in the Hamiltonian as
\begin{gather}
\sigma_i\left(\frac{p_\parallel^2}{2m}-\Delta\right) \xrightarrow{Hartree-Fock}\sigma_i\left(\frac{p_\parallel^2}{2m}-\Delta+Uc_d\right),
\label{eq1}
\end{gather}
where $p_\parallel$ is the direction of merging,  $\sigma_i$ is a Pauli matrix with $i=x$, $y$ or $z$, $m>0$ is the effective mass, $U$ is a local interaction.
The $c_d\gtrless 0$ is a non-universal constant coming from the Fock or Hartree terms, respectively, depending on whether an intersublattice  ($i=x$ or $y$ in Eq. \eqref{eq1}) or intrasublattice
($i=z$ in Eq. \eqref{eq1}) term drives
the merging. Its value depends only on  the details of the regularization and on
the spatial dimension $d$ of the model, in the $|\Delta|,|U|\ll W$ limit ($W$ the high energy cutoff).
$\Delta$ is the band-structure-dependent parameter that controls the energy difference between the saddle
points or the separation between the two Dirac cones. It measures in general the distance from the merging
transition, as
visualized in Fig. \ref{specevol}.

\begin{figure}[h!]
\includegraphics[]{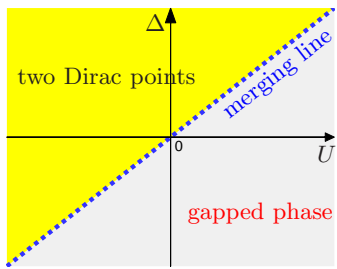}
\caption{The schematic phase diagram around the intersublattice induced merging transition of Dirac-Weyl points is shown on the $U-\Delta$ plane, 
without invoking any mean-field approximation. Along the merging line (blue dashed line)
with slope $c_d>0$,
the spectrum is quadratic along the merging direction, and linear in the others. For merging caused by intrasublattice processes, $c_d<0$, and the phase diagram is obtained by changing $U\rightarrow -U$.
}
\label{phasediag}
\end{figure}

In the presence of interactions, for $\Delta-Uc_d<0$, the system is gapped
and low energy excitations are absent, while $\Delta-Uc_d>0$ allows for low energy excitation around $p_\parallel=\pm\sqrt{2m(\Delta-Uc_d)}$ in the Brillouin zone. For $c_d>0$ (merging due to intersublattice terms), a
repulsive interaction moves the Dirac points towards each other and tries to open the gap, whereas an attraction
between particles in general tends to close the gap, and split the merged point into two Dirac cones.
This conclusion remains also true for a hole-like spectrum, $m<0$.
For $c_d<0$ (merging driven by intrasublattice terms), the above conclusions are reversed.
The evolution
of the spectrum and the $U-\Delta$ phase diagram are shown in Figs. \ref{specevol} and \ref{phasediag}.
Note that larger values of $\Delta$ and $U$ can renormalize the constant $c_d$, so that the strictly linear phase boundary holds true in the weak coupling limit in Fig. \ref{phasediag}.

The behavior described above can be probed in equilibrium by tuning the inter-particle interaction adiabatically
and measuring the resulting changes in the band structure. An alternative way to distinguish it from band
structure engineering is to focus on non-equilibrium dynamics. The constant $c_d$ depends in principle on the (non-equilibrium) particle occupation numbers,
 whereas when the splitting is determined by band structure engineering, $\Delta$ is obviously insensitive to it.
 
 Finally, we propose a 3D Hamiltonian quadratic in momentum, with isotropic dispersion, that generalizes the graphene bilayer Hamiltonian and describes two {\it coupled} Weyl points. We show that this Hamiltonian violates time reversal, and thus does not allow an opening of neither insulating, nor superconducting gaps. A weak interaction in this case causes either emergence of two Weyl points, or of the circle of the gapless points, depending on the interaction's sign and the details of the Hamiltonian. The analogous effect in the 2D bilayer graphene Hamiltonian is absent.

\section{ Merging and splitting of Dirac points}

\subsection{1D Dirac fermions}

\begin{figure}[h!]
\centering

\includegraphics[]{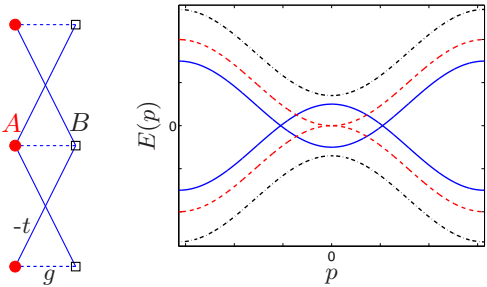}
\caption{A one-dimensional lattice with two atoms per unit cell, realizing two Dirac points for $g<2t$ (blue solid line), a quadratic band crossing for $g=2t$ (red dashed line)
and insulating behaviour for $g>2t$ (black dash-dotted lint) around half filling is shown. Here,
$g$ (dashed) and $-t$ (solid) are  the intra/intercell, intersublattice hopping integrals.}
\label{creutzlattice}
\end{figure}

The Hamiltonian, describing the system in Fig. \ref{creutzlattice}, is written in the basis of the two inequivalent atoms\cite{creutz} in the unit cell as
\begin{gather}
H=\sum_p\Psi^+_p\left[\begin{array}{cc}
0 & g-2t\cos(pa) \\
g-2t\cos(pa) & 0
\end{array}\right]\Psi_p,
\end{gather}
where $\Psi^+_p=(a^+_p,b^+_p)$, $a^+_p$ and $b^+_p$ creates fermions on sublattice $A$ and $B$, respectively.
 Its spectrum is $E_\pm(p)=\pm |g-2t\cos(pa)|$ with $a$ the lattice constant and $g,t>0$.
Interestingly, the lattice structure is reminiscent of a two-leg ladder after flipping every other bond in a given unit cell, 
although the spectrum is completely
different due to the distinct topology of the two lattices. For the sake of simplicity, we consider the limit of $g\simeq 2t$, where one can introduce the tuning parameter  $\Delta=2t-g$. After expanding the spectrum around the origin at $k=0$ we may focus on the low energy limit close to half filling, and  obtain the effective Hamiltonian  in first quantized form
\begin{gather}
H=\sigma_x \left(\frac{k^2}{2m}-\Delta\right),
\label{hamilton1d}
\end{gather}
where $m=1/2|t|a^2$ is the effective mass and $\sigma_x$ is a Pauli matrix, acting in the sublattice space.
The excitation spectrum of the Hamiltonian in Eq. \eqref{hamilton1d} is
\begin{gather}
E_\pm(p)=\pm \left|\frac{k^2}{2m}-\Delta\right|.
\end{gather}
Therefore, as long as $\Delta>0$, two Dirac points with linear dispersion around zeros of energy are present, 
while for $\Delta<0$, the system is a band insulator. The quadratic band crossing at $\Delta=0$ separates these two regions. This can be regarded as the one
dimensional version of Dirac point merging, as observed in Ref. \onlinecite{tarruell} in graphene-like systems. The density of states (DOS) at low energies then  changes from a constant to $\sim |E|^{-1/2}$, when $\Delta \rightarrow 0$.

We now consider the effect of a short-range electron-electron interaction on this system, which for spinless particles is written as
\begin{gather}
H_{int}=\frac{U}{N}\sum_{k,k',q} a^+_ka_{k+q}b^+_{k'}b_{k'-q}.\label{hint}
\end{gather}
This nearest-neighbor-interaction is the most relevant term in the cold atomic
setting, where short-range interactions are dominant.  $N$ is the number of
unit cells in the system. The interaction Hamiltonian can be decoupled at Hartree-Fock level as
\begin{gather}
H^{Fock}_{int}=-\frac UN \sum_{k,k'}\left[\langle b^+_{k'}a_{k'}\rangle_0 a^+_kb_k+\langle a^+_{k'}b_{k'}\rangle_0 b^+_k a_k\right],
\label{hinthf}
\end{gather}
where only the Fock term (with $k'=k+q$ from Eq. \eqref{hint}) is important,
the other terms arising from the decoupling being irrelevant for our discussion. 
\jav{The expectation value $\langle\dots\rangle_0$ is taken with respect to the non-interacting ground state, since we are considering only leading order perturbation theory. To the lowest order, the Hartree-Fock
diagrams contribute, and the validity of our approach is restricted to weak couplings, i.e. $|U/W|\ll 1$ with $W$ the high energy cutoff.
We emphasize that no self-consistency equations are to be satisfied since we do not perform any mean-field approximation.}

The resulting self energy then exhibits an off-diagonal component
\begin{gather}
\Sigma_k=\sigma_x Uc_1,
\end{gather} where
\begin{gather}
c_1=\frac{a}{2}\int\limits_{-k_c}^{k_c} \frac{dk}{2\pi}\textmd{sign}\left(\frac{k^2}{2m}-\Delta\right)
\approx \frac{a\sqrt{mW}}{\sqrt 2\pi },
\end{gather}
is a dimensionless positive constant in the $W\gg \Delta$ limit,  $W=k_c^2/2m$ being the high energy cutoff. As a result, the effective single-particle Hamiltonian changes into
\begin{gather}
H=\sigma_x \left(\frac{k^2}{2m}-\Delta+Uc_1\right).
\end{gather}
This means that the repulsive ($U>0$) interaction reduces $\Delta$, and tries to open up a gap in the system, while an attractive interaction,
$U<0$, favors splitting of the Dirac points.
Specifically, sitting right at the quadratic touching point with $\Delta=0$, the system can be driven to a gapped or a metallic (two Dirac
points) phase, depending on the repulsive or attractive nature
of the interaction. 
Had we taken a negative effective mass, it would not have changed the result, and the repulsive (attractive) interaction would again have driven the
system toward Dirac point merging (splitting). Identical conclusions can be obtained by retaining the original tight-binding spectrum, except when
$g=0$, when the off-diagonal component of the Hamiltonian averages to zero upon integration over the Brillouin zone.

\subsection{Merging Dirac points in 2D}

Similar phenomena are also  expected in two-dimensional systems, where Dirac points can be coupled in at least two distinct ways.

The first one is equivalent to  bilayer graphene\cite{evcastro}, when the linear band crossing at the Dirac cone changes to an isotropic quadratic one.
The density of states in this case becomes finite at zero energy, and in the presence of interactions the system becomes unstable to spontaneous breaking of some 
symmetry\cite{kaisun,vafek,cvetkovic}. We will call this case {\it coupling} of Dirac (or Weyl) points, and treat it separately in the sec. III, where we focus on the effects 
of interactions in the 3D generalization of the bilayer graphene.

The second option is the two dimensional generalization of Eq. \eqref{hamilton1d}, as analyzed in Refs. \onlinecite{tarruell,lim,montambaux},
where the merging of two Dirac points leads to anisotropic band crossing, linear in one direction and quadratic in the other. The density of states
in this case changes from linear, $\sim |E|$ to $\sim \sqrt{| E|}$. What happens in the presence of interactions is then not obvious,
 due to the peculiarity of the anisotropic band structure, which exhibits the graphene monolayer- and bilayer-like features in two different directions.

The effective Hamiltonian around the merging transition in 2D in the latter case is defined as \cite{montambaux}
\begin{gather}
H=\sigma_x \left(\frac{p_x^2}{2m}-\Delta\right)+\sigma_yvp_y,
\label{hamilton2d}
\end{gather}
and the interaction is assumed identical as in  Eqs. \eqref{hint} and \eqref{hinthf}, except, of course, with two-dimensional momenta.
The energy spectrum is  anisotropic,
\begin{gather}
E_\pm({\bf p})=\pm\sqrt{\left(\frac{p_x^2}{2m}-\Delta\right)^2+(vp_y)^2},
\end{gather}
and remains gapless for $\Delta>0$, with two Dirac cones, shows anisotropic band touching at $\Delta=0$\cite{pickett}, and finally yields insulating behavior for $\Delta<0$.

The self energy shows identical off-diagonal structure as in the 1D case: $\Sigma_k=\sigma_xUc_2$,  with
\begin{gather}
c_2=\frac{A_c}{2}\int  \frac{d^2p}{(2\pi)^2}\dfrac{\frac{p_x^2}{2m}-\Delta}{E_+({\bf p})}\approx W^{3/2} \frac{A_c\Gamma^2(3/4)}{3\pi^{5/2}}\frac{\sqrt m}{v},
\end{gather}
where $\Gamma(x)$ is the gamma function, $A_c$ is the area of the unit cell, and the final approximation is obtained by integrating over the $|E_\pm({\bf p})|<W$ modes centred on $p = 0$. Evidently,
again $c_2>0$, and as in the 1D case, a repulsive (attractive) interaction
favors Dirac point merging (splitting). Experimentally, the merging transition
has been observed in Ref. \onlinecite{tarruell} by tuning the hopping parameters of the underlying optical lattice.
Controlling the interaction, by tuning the Feshbach resonance or by modifying the lattice parameters for example,
seems straightforward, and therefore the experimental observation of the predicted interaction-induced merging and splitting appears possible. 
Similar effects have also been investigated in Refs. \onlinecite{volovikmerging,wang}.

\jav{The Dirac points in mono- and bilayer graphene sit at high symmetry points, so they are protected against moving in the Brillouin zone by interactions.
However, upon straining graphene, they can  be  moved in the Brillouin zone by a weak electron-electron interaction, although the merging transition would require entering 
into the strong coupling regime (by an interaction strength comparable to the bandwidth) due to the high energy barrier separating the Dirac cones, as was demonstrated in Ref. 
\onlinecite{wang}. 
In this case, our approach is not reliable any more and additional effects of the strong interaction need to be taken into account\cite{herbut,hjr}.}

Let us note that the above conclusions are valid when the merging occurs due to interspecies or intersublattice terms (i.e. $\sigma_x$ or $\sigma_y$ terms in the Hamiltonian), in which case
the intersublattice hopping integrals drive the merging  and the Fock term from Eq. \eqref{hinthf} dominates the interaction.
There are also lattice models, which contain $\sigma_z$ instead of  $\sigma_x$ in Eq. \eqref{hamilton2d}.
For example, a square lattice in Ref. \onlinecite{diracsquare} with identical hoppings $t$ along the chains in the $x$ direction
and alternating hoppings $\pm t$ for even/odd  chains in the $y$ direction host an even number of Dirac cones. Adding an opposite on-site potential to
the two sublattices will move the cones towards each other, and will eventually cause their merging.
In this case of merging, driven by the intrasublattice terms,
the Hartree decoupling (with $q=0$ in Eq. \eqref{hint}) of the interaction as
\begin{gather}
H^{Hartree}_{int}=\frac UN \sum_{k,k'}\left[\langle b^+_{k'}b_{k'}\rangle_0 a^+_ka_k+\langle a^+_{k'}a_{k'}\rangle_0 b^+_k b_k\right]
\label{hinthfz}
\end{gather}
is appropriate instead of the Fock one in Eq. \eqref{hinthf}, i.e. the minus sign is lacking due to the fermionic nature of the excitations.
Therefore, the role of the interaction is reversed in this case, namely repulsive/attractive interaction leads to Weyl point splitting/gap opening.

\subsection{Merging Weyl points in 3D}

Inspired by the exciting  physics of graphene and topological insulators, nodal semimetals in 3D are currently under investigation. \cite{balents, burkov, delplace} The Weyl Hamiltonian exhausts all three Pauli matrices, and is written as
\begin{gather}
H=v_F\left(\sigma_xp_x+\sigma_yp_y+\sigma_zp_z\right).
\end{gather}
Opening a gap in the above Hamiltonian is obviously impossible, since coupling anything to one of the Pauli matrices only shifts the position of the zero energy state
in the momentum space.  Nevertheless, two such Weyl points can be manipulated similarly to the previously discussed lower dimensional cases. The effective Hamiltonian,
describing two Weyl points near their merging transition is\cite{burkov, delplace}
\begin{gather}
H=\sigma_x \left(\frac{p_x^2}{2m}-\Delta\right)+\sigma_yv_yp_y+\sigma_zv_zp_z,
\label{hamilton3d}
\end{gather}
with the energy spectrum
\begin{gather}
E_\pm({\bf p})=\pm\sqrt{\left(\frac{p_x^2}{2m}-\Delta\right)^2+(v_yp_y)^2+(v_zp_z)^2}.
\end{gather}
In the region with two Weyl points, the energy increases linearly with momentum in the low energy limit, and the DOS exhibits $\sim E^2$ characteristics. This  changes to a DOS $\sim |E|^{3/2}$ at the merging point, when the spectrum
becomes strongly anisotropic.
The self energy due to the interaction is again $\Sigma_k=\sigma_xUc_3$, with the constant
\begin{gather}
c_3=\frac{V_c}{2}\int  \frac{d^3p}{(2\pi)^3}\dfrac{\frac{p_x^2}{2m}-\Delta}{E_+({\bf p})}\approx W^{5/2}\frac{V_c\sqrt{m}}{15\sqrt 2  \pi^2 v_yv_z} > 0,
\end{gather}
where $V_c$ is the volume of the unit cell. Similarly to the lower-dimensional cases already discussed, the interaction can be used to merge or split
Weyl points in 3D in a qualitatively similar way.

In case of merging by the $\sigma_z$ term \cite{balents, burkov, delplace} as opposed to the $\sigma_x$ term in Eq. \eqref{hamilton3d} or $\sigma_y$, the above conclusions are reversed and
 the repulsive (attractive) interaction would again drive the
system toward Dirac point splitting (merging).

\section{Coupling Weyl points \`a la bilayer graphene}

\subsection{3D}

In 3D, however, there exist another yet apparently unexplored possibility, that can be conveniently understood as a 3D version of  the  bilayer graphene, with two Weyl points coupled. The resulting Hamiltonian, describing the low energy modes,  is analogous to that in bilayer graphene: 
\begin{gather}
H=\frac{1}{2m}\left(\sigma_x 2p_xp_z+ \sigma_y 2p_yp_z+\sigma_z\left[p_z^2-p_\perp^2\right]\right),
\label{hamilton4}
\end{gather}
with an isotropic spectrum that reads
\begin{gather}
E_\pm({\bf p})=\pm \frac{|{\bf p}|^2}{2m}.
\end{gather}
 Here $p_\perp^2=p_x^2+p_y^2$, and we take $m > 0$, for simplicity.  The DOS now varies as $\sqrt{|E|}$. Such a quadratic-Weyl Hamiltonian is also stable to opening of the (insulating) gap, since it uses up all three of the Pauli matrices. In fact, it is even more stable than the usual Weyl Hamiltonian in Eq. (13): since it breaks the time-reversal symmetry, even the opening of the superconducting gap is no longer possible.\cite{herbutPRD} To see this, recall that the operator for the time-reversal is {\it antilinear}: $I_t = U K$, where $U$ is unitary, and $K$ is the complex conjugation. The unitary operator $U$ would therefore have to commute with the two real Pauli matrices $\sigma_1$ and $\sigma_3$, and anticommmute with the imaginary $\sigma_2$, in order for the Hamiltonian (18) to be invariant under time reversal. Such a two-dimensional matrix $U$ obviously does not exist. Since the standard s-wave superconducting gap is nothing but the Cooper pairing between the time-reversed states of the Hamiltonian, which are now non-existent, in this case the superconducting gap becomes equally (algebraically) impossible as the usual insulating gap.\cite{herbutPRD}

Nevertheless, by treating the interaction at the lowest (Hartree-Fock) level, we again obtain the self-energy similar to the previous examples, using the Hartree decoupling in Eq.
\eqref{hinthfz} as
\begin{gather}
\Sigma_k= \sigma_zU\tilde c,
\end{gather}
where
\begin{gather}
\tilde c=\frac{V_c}{2}\int  \frac{d^3p}{(2\pi)^3}\dfrac{p_\perp  ^2 - p_z^2 }{2mE_+({\bf p})}\approx  \frac{(2mW)^{3/2}V_c }{36\pi^2} > 0.
\label{tildec}
\end{gather}

As a result, to leading order the effective single-particle Hamiltonian in Eq. \eqref{hamilton4} changes with the inclusion of the  self-energy, and the
corresponding spectrum becomes
\begin{gather}
E_\pm({\bf p})=\pm\frac{\sqrt{4p_z^2p_\perp^2+(2mU\tilde c+p_z^2-p_\perp^2)^2}}{2m}.
\end{gather}
Due to the presence of interaction the nature of the excitations of the system is now modified in the following way:
first, the gapless point in the non-interacting system is located at ${\bf p}=0$. Second, for a repulsive ($U>0$) interaction,
the location of the gapless excitations is the circle  at $p_x^2+p_y^2=2mU\tilde c$ in the $p_z=0$ plane, causing the DOS to increase with $|E|$ away from the band touching,
while third, an attractive ($U<0$) interaction leads to two Weyl points with linear dispersion at $p_x=p_y=0$ and $p_z=\pm \sqrt{2m|U|\tilde c}$,
and consequently the DOS varies as $E^2$ at low energies.

Had we coupled the $p_z^2-p_\perp^2$ term to $\sigma_{x}$ or $\sigma_y$, the Fock term from of Eq. \eqref{hinthf} would reverse
the effect of the interaction, similarly to how it occurs at the merging transition of Dirac-Weyl points.

\subsection{Lower dimensions}

 Taking one of the $x$ and $y$ components of the momentum to zero in Eq. (18) reduces the Hamiltonian to the case of the quadratic band crossing in the AB-stacked bilayer graphene. 
The term analogous to Eq. (21) then cancels out in bilayer graphene, where only the combination $p_x^2-p_y^2$ would appear in the numerator of Eq. \eqref{tildec}, and would thus yield zero after the integration. 
 
 The 1D case may also be obtained by setting $p_y=p_z=0$, when Eq. \eqref{hamilton4} reduces to the
same effective Hamiltonian as in Eq. \eqref{hamilton1d}, though with negative effective mass.

\section{Summary}

We have studied the effect of inter-particle interaction to leading order perturbation theory (without resorting to any kind of mean-field approximation),
when two Dirac-Weyl points are merged along certain direction, or coupled like
in bilayer graphene. In the former case, and in all physical dimensions, a repulsive interaction favors the merging of the Dirac points
and the eventual opening of the gap, whereas an attraction pushes the Dirac points apart, regardless to the sign of the effective mass
along the merging direction, when the merging point is reached by tuning the intersublattice terms in the Hamiltonian.
In case of intrasublattice term driven merging, the above conclusions are reversed.

We can also envisage an interesting scenario of reaching the merging point by simultaneously tuning inter- and intrasublattice terms, in which case
a delicate competition between these two processes and the sign of the interaction would decide on whether a gap opens or two Dirac points appear.

The latter case of coupling of the Dirac points
in 1D is identical to merging, while the Hartree-Fock corrections are harmless when Dirac points are coupled in 2D. The coupling of the Weyl
point in 3D, however, changes the topology of the low energy excitations, which remain gapless to lowest order in the interaction, but change
from a circle of gapless points, to two separate Weyl points in momentum space.

\jav{These band structures arise not from any kind of symmetry breaking, but rather follow from single particle band structure renormalization by interaction effects.
We therefore expect them to be robust in any dimension.}

\begin{acknowledgments}

BD has been  supported by the Hungarian Scientific  Research Funds Nos. K101244, K105149, CNK80991, by the Bolyai Program of the Hungarian Academy of Sciences and partially by the ERC Grant Nr. ERC-259374-Sylo. IFH has been supported by the NSERC of Canada.
\end{acknowledgments}

\bibliographystyle{apsrev}

\bibliography{refgraph}

\end{document}